\documentstyle[prl,multicol,aps]{revtex}
\def\twobeone{\end{multicols}
\vskip.6pc
\noindent
\vrule width3.375in height.2pt depth.2pt \vrule depth0em height1em\hfill
\vskip.6pc}

\def\onebetwo{
\vskip.6pc
\indent
 \hfill\vrule depth1em height0pt \vrule width3.375in height.2pt depth.2pt
\vskip.6pc
\begin{multicols}{2}\noindent}

\begin{document}

\draft

%\preprint{ZIMP-99-}

\title
{ Thermodynamics of Kondo model with electronic interactions}

\author
{Shi-Jian Gu$\dagger$, You-Quan Li$\dagger$ and Hai-Qing Lin$\ddagger$} 

\address
{
$\dagger$Zhejiang Institute of Modern Physics,
Zhejiang University, Hangzhou 310027, P. R. China\\
$\ddagger$Department of Physics, Chinese University of Hong Kong, 
P. R. China 
}

\date{Received: Feb. 15, 1999} 

\maketitle

\begin{abstract}

On the basis of Bethe ansatz solution of 
one dimensional Kondo model
with electronic interaction,  
the thermodynamics equilibrium of the system in finite temperature
is studied in terms of
the strategy of Yang and Yang. 
The string hypothesis in the spin rapidity is discussed extensively. 
The thermodynamics quantities, such as specific heat and magnetic
susceptibility, are obtained.
\end{abstract}


%\twocolumn
\begin{multicols}{2}
\section{Introduction}

It is known that the study of exact solutions is 
helpful for the understanding of non-perturbative effects in the 
strongly correlated electronic systems. 
The exact solution of the one dimensional 
Kondo\cite{Kondo} model with linearized dispersion 
in the absence of electronic  
interaction was found in \cite{Andrei,Wiegmann}.
The model with quadratic dispersion in the presence of electronic
was shown to be exactly solvable at some value of 
electron-impurity coupling\cite{LiB}.

In present paper, we study the thermodynamics of Kondo model 
in the presence of the electronic interactions.
The general thermal equilibrium are discussed exactly
on the basis of the known Bethe ansatz solutions of the model\cite{LiB}.
The specific heat and magnetic susceptibility are  obtained analytically
in general and given  explicitly in strong-coupling limit.
The specific heat and the magnetic susceptibility
at low temperature are discussed.
Next section we briefly exhibit  the model under consideration and its
Bethe ansatz solution. In section III, we demonstrate the string hypothesis
and write out the Bethe ansatz equation in the presence of complex roots.
In section IV, we consider the thermodynamics limit by introducing 
the density of roots and holes. In section V we derive 
the free energy of the system at thermal equilibrium 
according to the strategy of Yang and Yang\cite{YYang}.
In section VI, We calculate the thermodynamics quantities, such 
as specific heat and magnetic susceptibility.
In the case of strong coupling limit, we
find that the  contributions of both electrons and impurity to 
the specific heat
and the magnetic susceptibility is Fermi-liquid like.

\section{The model and its spectrum}

The model Hamiltonian of a correlated electronic system  
which we consider reads
\begin{eqnarray}
H_0=\sum_{\bf k}\varepsilon({\bf k})C^{\ast}_{{\bf k}a}C_{{\bf k}a} 
       \hspace{36mm}\nonumber\\
   +\sum_{ {\bf k}_1,{\bf k}_2, {\bf k}_3,{\bf k}_4 }
    u\delta({\bf k}_1+{\bf k}_2 ,{\bf k}_3+{\bf k}_4) 
         C^{\ast}_{ {\bf k}_4 a} C^{\ast}_{ {\bf k}_3 a} 
         C_{ {\bf k}_2 a} C_{ {\bf k}_1 a},\nonumber  
\end{eqnarray}      
where $C_{ {\bf k} a }$ annihilates an electron with 
momentum ${\bf k}$ and spin component $a$, and 
$\varepsilon(k)= k^{2}/2 $ (in units of $\hbar$ and of the electron mass).
The electrons are coupled by
both spin and  charge interactions to a localized impurity,
\[ 
H_I=J\Psi^{\ast}_a(0){\bf S}_{ab}\Psi_{b}(0)\cdot{\bf S}^0 
     + V\Psi^{\ast}_a (0)\Psi_a (0),
\]
where the field $\Psi_a$ is the Fourier transform of $C_{{\bf k}a}$,
${\bf S}^0$ is the spin of the impurity and  
${\bf S}$ is the spin  of the electrons in the  band.

The present model was solved
exactly by Bethe ansatz with periodic boundary conditions\cite{LiB},
It was shown that the model is integrable when the correlation strength 
$u$ is proportional to the $J$, the strength of 
electron-impurity coupling via spin.  
The  Bethe Ansatz equations
for the spectrum are 
\twobeone
\begin{eqnarray}
e^{-ik_j L }= 
       e^{-i\theta( k_j )}\prod_{\nu=1}^{M}
          \frac{\lambda_{\nu} - k_j + iu/2}
               {\lambda_{\nu} - k_j - iu/2},
                    \hspace{16mm}\nonumber \\
 -\prod_{\nu=1}^{M}\frac{\lambda_{\nu}-\lambda_{\mu}+iu}
                        {\lambda_{\nu}-\lambda_{\mu}-iu}
 =\frac{\lambda_{\mu}-iu/2}
        {\lambda_{\mu}+iu/2}
   \prod_{l=1}^{N}\frac{\lambda_{\mu}-k_l-iu/2}
                       {\lambda_{\mu}-k_l+iu/2},
\label{eq:BAE}
\end{eqnarray}
\onebetwo  
where $\theta(k_j)=2\tan^{-1}(k_j/u)$.
In the approximation $k_j \sim k_l$ for any $j, l$, 
the $S$ matrix of electron-electron will be independent of 
$u$. Then the Yang-Baxter equation will give no restrictions 
on between $u$ and $J$. This makes it easy to understand
the usual Kondo problem where the linear dispersion
relation is adopted, whence the model is solvable at any value of $J$. 

\section{String hypothesis}

For the ground state (i.e. at zero temperature), the $k'$s and $\lambda'$s
are real roots of the Bethe ansatz equation (\ref{eq:BAE}). For the excitated
state (i.e. at nonzero temperature), however, they can be complex 
roots\cite{Woynarovich,Takahashi}. 
We will not take account of the complex roots in the 
charge sector $k$ for repulsive interaction since it
does not happen at low temperature. The complex roots $\lambda$ in spin sector
always for a ``bound state'' with several other $\lambda'$s, which arises
from the consistency of both hand side of 
the Bethe ansatz equation\cite{Woynarovich}. 
The complex roots with the same real part $\lambda^n_\beta$ form a 
$n$-string,
\begin{eqnarray}
\Lambda_\beta^{n m}=\lambda_\beta^n + i\frac{u}{2}m + O(\exp(-\delta N)), 
  \,\,\, (\delta > 0) \nonumber\\
m=-n+1, -n+3, \cdots, n-3, n-1.
\end{eqnarray}
The set of roots $\{ \lambda_\nu | \nu =1, 2, \cdots, M\}$
is then partitioned into a set of $n$-strings
$\{ \Lambda^n_\beta |m=-n+1, -n+3, \cdots, n-3, n-1;
    \beta= 1, 2, \cdots, M_n \}$.
Obviously, 
$$ M=\sum_{n=1}^{\infty}n M_n ,$$
where $M_n$ denotes the number of $n$-strings.

Substituting those $n$-strings into the eq.(\ref{eq:BAE}), we can find that the 
product of the fractions for the roots within the same $n$-string
reduce to 
$(\lambda^n_\beta - k_j +i n u/2)/(\lambda^n_\beta -k_j - i n u/2)$
because of the alternative elimination between denominator and numerator.
Hence the Bethe ansatz equation (\ref{eq:BAE}) becomes
\twobeone
\begin{equation}
e^{-ik_jL}=e^{-i\theta(k_j)}
    \prod_{\beta n}\frac{\lambda_\beta^n-k_j+inu/2}
                        {\lambda_\beta^n-k_j-inu/2},
\label{eq:BAEcharge}
\end{equation}
and
\begin{equation}
-\prod_{\beta m}\prod_{q=-m+1}^{m-1}
      \frac{\Lambda_\beta^{m q}-\Lambda_\alpha^{n p}+iu}
           {\Lambda_\beta^{m q}-\Lambda_\alpha^{n p}-iu}
 =\frac{\Lambda_\alpha^{n p}-iu/2}
       {\Lambda_\alpha^{n p}+iu/2}
   \prod_{l=1}^N
     \frac{\Lambda_\alpha^{n p}-k_l-iu/2}
          {\Lambda_\alpha^{n p}-k_l+iu/2}.
\label{eq:BAEspin}
\end{equation}
The product of those equations (\ref{eq:BAEspin}) for
$p= -n+1, -n+3, \cdots, n-3, n-1$ gives rise to
\begin{eqnarray}
(-1)^n \prod_{\beta m}
  \frac{\lambda_\beta^m-\lambda_\alpha^n+i(m+n)u/2}
       {\lambda_\beta^m-\lambda_\alpha^n-i(m+n)u/2} \times\nonumber\\
  \left[
    \frac{\lambda_\beta^m-\lambda_\alpha^n+i(m+n-2)u/2}
         {\lambda_\beta^m-\lambda_\alpha^n-i(m+n-2)u/2}
    \cdots\frac{\lambda_\beta^m-\lambda_\alpha^n+i(|m-n|+2)u/2}
               {\lambda_\beta^m-\lambda_\alpha^n-i(|m-n|+2)u/2}
  \right]^2 \times\nonumber\\
  \frac{\lambda_\beta^m-\lambda_\alpha^n+i|m-n|u/2}
             {\lambda_\beta^m-\lambda_\alpha^n-i|m-n|u/2}
=\frac{\lambda_\alpha^n-inu/2}
      {\lambda_\alpha^n+inu/2}
    \prod_l^N\frac{\lambda_\alpha^n-k_l-inu/2}
                  {\lambda_\alpha^n-k_l+inu/2}.
\label{eq:BAEstring}
\end{eqnarray}
Taking the logarithm of eq.(\ref{eq:BAEcharge}) and eq.(\ref{eq:BAEstring})
we have
\begin{eqnarray}
k_j=\frac{2\pi}{L}I_j + \frac{1}{L}\theta(k_j)
     +\frac{1}{L}\sum_{\beta n}\Theta_{n/2}(\lambda_\beta^n-k_j),
                 \nonumber\\
\Theta_{n/2}(\lambda_\alpha^n)
  +\sum_{l=1}^N\Theta_{n/2}(\lambda_\alpha^n-k_l)
=2\pi J_\alpha^n
  -\sum_{\beta m p}A_{nmp}\Theta_{p/2}(\lambda_\beta^m-\lambda_\alpha^n),
\label{eq:secular}
\end{eqnarray}
where $\Theta_\rho (x)=2\tan^{-1}(\frac{x}{\rho u})$ and
\[
A_{nmp}=\left\{\begin{array}{ll}
1, & {\rm for}\; p=m+n, |m-n|(\neq 0),\\
2, & {\rm for}\; p=n+m-2, n+m-4,\cdots,|n-n|+2,\\
0,  & {\rm  otherwise.}
\end{array}\right.
\]
\onebetwo
The $I_j$ and $J_\alpha^n$ are quantum numbers,
the $I_j$  are integers or half-integers depending on whether
$M$ is even or odd,
the $J^n_\alpha$ are integers or half-integers depending on whether
$N-M_n -n+1$ is even or odd. 

\section{The thermodynamics limit}

The trasendental equations (\ref{eq:secular}) for the real parts 
of the complex roots are
obviously difficult to solve. It will be convenient to consider the
thermodynamics limit\cite{Lowenstein}, 
i.e. $N\rightarrow\infty$, $L\rightarrow\infty$ but
$D=N/L$ is fixed. 
Introducing the density distribution of roots and holes 
\begin{eqnarray}
\frac{1}{L}\frac{d I(k)}{dk} &=& 
          \rho(k) + \rho^h(k), \nonumber\\
\frac{1}{L}\frac{d J^n(\lambda)}{d\lambda} &=&
            \sigma_n(\lambda)+\sigma_n^h(\lambda),
\label{eq:densities}
\end{eqnarray}
we obtain from (\ref{eq:secular}) the following set of integral equations,
\twobeone
\begin{eqnarray}
\rho(k)+\rho^h(k) &=& \frac{1}{2\pi}
     -\frac{1}{L}K_1(k)
      +\sum_{n=1}^\infty K_{n/2}(k|\lambda')\sigma(\lambda'),\nonumber\\
\sigma_n(\lambda)+\sigma_n^h(\lambda) &=& \frac{1}{L}K_{n/2}(\lambda)
     +K_{n/2}(\lambda|k')\rho(k')
      -\sum_{m p}A_{nmp}K_{p/2}(\lambda|\lambda')\sigma_m(\lambda'),
\label{eq:integrals}
\end{eqnarray}
\onebetwo
where $K_n(x)=\pi^{-1}nu/(n^2u^2 +x^2)$. We have adopted
a notation convention
$K_n(x|y)\rho(y)=\int K_n (x-y)\rho(y)dy$ etc. in the above.

In term of the density distributions, 
the energy and the concentration of electrons 
as well as the number of down spins are given by 
\begin{eqnarray}
E &=& L\int_{-\infty}^\infty dk\rho(k)k^2, \nonumber\\
D &=& \frac{N}{L}=\int^{\infty}_{-\infty} 
         dk\rho(k), \nonumber \\
\frac{M}{L} &=& \sum_{n=1}^\infty n\int^\infty_{-\infty} 
         d\lambda\sigma_n(\lambda).\nonumber 
\end{eqnarray} 
Thus the magnetic moment of the system is,
\begin{eqnarray}
{\cal M} &=& \frac{1}{2}(N-2M) + S^z_{imp} \\
  \,&=& \frac{L}{2}\int_{-\infty}^\infty\rho(k)dk 
         -L\sum_{n=1}^\infty n\int_{-\infty}^\infty
            \sigma_n(\lambda)d\lambda + S^z_{imp},
\label{eq:totalspin}
\end{eqnarray}
where $S^z_{imp}$ stands for the spin of the impurity,
and the $g$ factor is put to unit.

The ground state of the present model is a Fermi
sea described by
$\rho(k)$ with real rapidity $\lambda$, i.e.
$\rho(k)=0$ for $|k|>k_F$ and $\rho^h(k)=0$ for $|k|<k_F$;
$\sigma_1(\lambda)\neq 0$ but 
$\sigma_n(\lambda)=0$ $(n>1)$,
which is the case at zero temperature. Away from zero temperature,
The density distributions of roots  and holes with respective
to the  momentum $k$
and the spin rapidity $\lambda$ 
should be determined by the principles of statistical physics. Next
section we will discuss this issue extensively on the basis of 
the strategy of Yang and Yang\cite{YYang}.

\section{Thermal equilibrium} 

For a given $\rho(k)$ and $\rho^h(k)$, the number of roots and that
of holes in the neighborhood $dk$ are 
$L\rho dk$ and $L\rho^h dk$ respectively. Obviously, the total number
of roots and holes in the neighborhood is $L(\rho +\rho^h)dk$. 
For a given 
$\sigma_n(\lambda)$ and $\sigma_n^h(\lambda)$,
$L\sigma_n d\lambda$  and
$L\sigma_n^h d\lambda$ 
give rise to the number of $n$-string and the number of the vacancies 
of $n$-strings (holes) in the neighborhood $d\lambda$, while  
$L(\sigma_n +\sigma^h_n)d\lambda$ gives rise to the total number of 
$n$-string and vacancies of $n$-string. Thus the total number of 
the possible choice of state in $dk d\lambda$ being consistent with given
distribution functions in both charge and spin sectors is
\[
\Xi (k,\lambda)= 
   \frac{[L(\rho+\rho^h)dk]!}
        {[L\rho{dk}]![L\rho^h dk]!}
     \prod_n\frac{[L(\sigma_n+\sigma_n^h)d\lambda]!}
                 {[L\sigma_n d\lambda]![L\sigma_n^hd\lambda]!}.
\]
As the total number of all possible state for given distribution functions
is 
\[
\Xi =\prod_{k \lambda}\Xi(k,\lambda),
\]
the total entropy $S$ will be obtained by taking logarithm of $\Xi$,
\twobeone
\begin{eqnarray}
S/L=\int\left\{ [\rho(k)+\rho^h(k)]\ln [\rho(k)+\rho^h(k)]
              -\rho(k)\ln\rho(k)-\rho^h(k) \ln \rho^h(k)
       \right\} dk \nonumber\\
  +\sum_n\int \left\{ 
                   [\sigma_n(\lambda)+\sigma_n^h(\lambda) ]
                   \ln[\sigma_n(\lambda)+\sigma_n^h(\lambda)]
        -\sigma_n(\lambda)\ln\sigma_n(\lambda)
                 -\sigma_n^h(\lambda)\ln\sigma_n^h(\lambda)
            \right\} d\lambda,
\label{eq:entropy}
\end{eqnarray}
\onebetwo
where the Boltzmann constant is put to unit. 

In the presence of the external magnetic field $H$, we must 
add a Zeeman term to the original Hamiltonian. As the Zeeman term
commute with the original Hamiltonian, the Bethe ansatz solution is still
valid for present case. 
Therefore the energy of the system in the presence of external magnetic 
field will be  
\begin{equation}
E/L=\int(k^2- H)\rho(k)dk+
    \sum_{n=1}^{\infty}2n H\int\sigma_n(k)dk,
\label{eq:energy}
\end{equation}

In order to obtain the thermal equilibrium at temperature $T$, 
we should maximize the the contribution  
to partition function from
the state described by the density distributions of roots and holes.
As maximizing the partition function is equivalent 
to minimizing the free energy
$F=E-TS-\mu N$.
Here $S$ and $E$ are given by 
eq.(\ref{eq:entropy}) and eq.(\ref{eq:energy}),
$\mu$ is the chemical potential for canonical ensembles. 
The $\mu$ plays the role of the Lagrangian multiplier for the condition
$L\int\rho(k)dk=N=constant$ if one minimizing the Helmholtz free energy
$\Omega =E-TS$. This constranit implies that the essemble has fixed number
of partiles. A constraint that the number of down spins are fixed was imposed
in ref.\cite{Lai} when discussing delta Fermi gas, 
whereas we will not impose no physics constraints in the 
following discussion.

Making use of 
the relations derived from eq.(\ref{eq:secular})
\twobeone
\begin{eqnarray}
\delta\rho^h(k) &=&
   -\delta\rho(k)+\sum_n K_{n/2}(k|\lambda)\delta\sigma_n(\lambda),
                    \nonumber\\
\delta\sigma_n^h(\lambda) &=& 
   -\delta\sigma_n(\lambda)+K_{n/2}(\lambda|k)\delta\rho(k)
     -\sum_{mp}A_{mnp}K_{p/2}(\lambda|\lambda')\delta\sigma_m(\lambda'),
\label{eq:delta}
\end{eqnarray}
we obtain the following conditions from  the minimum condition 
$\delta F=0$, namely
\begin{eqnarray}
\epsilon(k) &=& 
      -\mu+k^2-H-T\sum_n K_{n/2}(k|\lambda)\ln(1+e^{-\xi_n(\lambda)/T}),
                    \label{eq:epsilon}\\
\xi_n(\lambda) &=&
      2n H-T K_{n/2}(\lambda|k)\ln(1+e^{-\epsilon(k)/T})\nonumber\\
   \,&\,&     +T\sum_{mp}A_{nmp}K_{p/2}(\lambda|\lambda')
                   \ln(1+e^{-\xi_m(\lambda')/T}),
\label{eq:xi}
\end{eqnarray}
\onebetwo
where we have written 
\begin{eqnarray}
\frac{\rho^h(k)}{\rho(k)}=\exp[ \epsilon(k)/T ],
         \nonumber\\
\frac{\sigma_n^h(\lambda)}{\sigma_n(\lambda)}
       =\exp[ \xi_n(\lambda)/T ].
            \nonumber
\end{eqnarray}

Principally, once $\epsilon(k)$ and $\xi(\lambda)$ are solved from 
eq.(\ref{eq:epsilon}-\ref{eq:xi}), the equilibrium distributions $\rho(k)$
and $\sigma_n(\lambda)$ at temperature $T$ will be known from 
the following relations,
\twobeone
\begin{eqnarray}
\rho(k)[1 + e^{\epsilon(k)/T}]
&=&\frac{1}{2\pi}-\frac{1}{L}K_1(k) + \sum_n K_{n/2}(k|\lambda')
                 \sigma_n(\lambda'), \nonumber\\
\sigma_n(\lambda)[1+e^{\xi(\lambda)/T}]
&=&\frac{1}{L}K_{n/2}(\lambda) + K_{n/2}(\lambda|k')\rho(k')\nonumber\\
 \,&\,& -\sum_{m q}A_{nmq}K_{q/2}(\lambda|\lambda')\sigma_m(\lambda').
\label{eq:distributionatT}
\end{eqnarray}
The free energy per unit length reads 
\begin{eqnarray}
 F/L = \int dk\rho(k)\left[
          k^2 -\epsilon(k) - H - T(1+e^{\epsilon(k)/T})
           \ln(1+e^{-\epsilon(k)/T})
                     \right]    \nonumber\\
 +\sum_n \int d\lambda\sigma_n(\lambda)\left[
         2nH -\xi_n(\lambda) - T(1+e^{\xi_n(\lambda)/T})
           \ln(1+e^{-\xi_n(\lambda)/T}) \right]. 
\label{eq:fenergy}
\end{eqnarray}

Integrating eq.(\ref{eq:epsilon}) over $k$ after multiplying it 
with $D^{-1}\rho$, we get the chemical potential
\begin{eqnarray}
\mu &=&\frac{1}{D}\int(k^2-\epsilon(k)-H)\rho(k){dk}\nonumber\\
 \,&\,&  -\frac{T}{D}\sum_n\int\int K_{n/2}(k-\lambda)
    \ln(1+e^{-\xi_n(\lambda)/T})\rho(k)d\lambda dk.
\label{eq:intrho}
\end{eqnarray}
Integrating eq.(\ref{eq:xi}) over $\lambda$ and summing over $n$
after multiplying it with $D^{-1}\sigma_n$, 
we obtain the following relation
\begin{eqnarray}
\sum_n\int\xi_n(\lambda)\sigma_n(\lambda)d\lambda = 
      \sum_n 2n H\int\sigma_n(\lambda)d\lambda   
          \hspace{2cm}   \nonumber\\
 -T\sum_n\int\int K_{n/2}(\lambda-k)\ln(1+e^{-\epsilon(k)/T})
     \sigma_n(\lambda)dk d\lambda
          \hspace{1cm}    \nonumber\\
 +T\sum_{mnq}A_{mnq}\int\int K_{q/2}(\lambda-\lambda')
     \ln(1+e^{-\xi_m(\lambda')/T})\sigma_n(\lambda)d\lambda'd\lambda.
\label{eq:intxi}
\end{eqnarray}
Using the relations (\ref{eq:intrho}-\ref{eq:intxi}), we can
write out the free energy in terms of $\epsilon$ and $\xi$ only, 
\onebetwo
\begin{eqnarray}
F=\mu N+T\int (K_1(k)-\frac{L}{2\pi})
          \ln(1+e^{-\epsilon(k)/T})dk
             \nonumber\\
     -T\sum_n\int K_{n/2}(\lambda)
        \ln(1+e^{-\xi_n(\lambda)/T})d\lambda.
\label{eq:ffenergy}
\end{eqnarray}
Consequently, the partition function is given by
\begin{equation}
Z=e^{-F/T},
\label{eq:pfunction}
\end{equation}
where the Boltzmann constant is put to unit.

The thermodynamics functions,
partition function $Z$,
free energy $F$, and thermal potential $\Omega$ etc, are of 
importance, as knowing either of them, one is able to calculate all  
thermodynamics properties of the system in principle. 
However, it is difficult to 
obtain an analytic expressions of $\epsilon(k)$ and
$\xi_n(\lambda)$ 
from the coupled non-linear integral equations (\ref{eq:epsilon}) and
(\ref{eq:xi}).
So we are not able to derive explicit results for thermodynamics quantities.
Moreover, 
we will obtain some plausible results for some special cases in 
next section.

\section{Thermodynamic quantities}

In general, the free energy of our model should be calculated by formula
(\ref{eq:ffenergy}), where the $\epsilon(k)$ and $\xi(\lambda)$ are 
determined by eq.(\ref{eq:epsilon}-\ref{eq:xi}). Then the other 
thermodynamic quantities are obtainable from thermodynamic relations.

In the Appendix, we show that if $\mu$, $\epsilon$ and $\xi_n$ are
implicit functions of some thermodynamic quantities $x$ (such as $T$, $L$), 
the derivative
of eq.(\ref{eq:ffenergy}) with respect to $x$ is the same as
the partial derivative of eq.(\ref{eq:ffenergy}) with respect to
the explicit variable $x$. 
It is easy to get the pressure of the system
\begin{equation}
P=-\frac{\partial F}{\partial L}
  = \frac{T}{2\pi}\int\ln(1+e^{-\epsilon(k)/T})dk,
\label{eq:442}
\end{equation}
which is the  same as Yang and Yang's expression formally. However,
the integral equation which $\epsilon (k)$ obeys are different. 
In our present case
the contributions from both impurity and the spin rapidity are involved.

In terms of $\epsilon$ and $\xi$,
the entropy $S=-(\partial F/\partial T)$ becomes 
\begin{eqnarray}
S &=& -\int\ln(1+e^{-\epsilon(k)/T})\left[
        -\frac{L}{2\pi}+\frac{1}{\pi}
           \frac{u}{u^2+k^2}    \right]dk   \nonumber\\
\,&\,& -\int\frac{e^{-\epsilon(k)/T}}{1+e^{-\epsilon(k)/T}}
         \frac{\epsilon(k)}{T} \left[
          -\frac{L}{2\pi}+\frac{1}{\pi}
             \frac{u}{u^2+k^2} \right]dk     \nonumber\\
 \,&\,& +\sum_n\int\frac{1}{\pi}\frac{nu/2}{(nu/2)^2+\lambda^2}
     \ln(1+e^{-\xi_n(\lambda)/T})d\lambda    \nonumber\\
 \,&\,& +\sum_n\int\frac{1}{\pi}\frac{nu/2}{(nu/2)^2+\lambda^2}
    \frac{\xi_n(\lambda)/T}{1+e^{\xi_n(\lambda)/T}}d\lambda.
\label{eq:theentropy}
\end{eqnarray}
The other thermodynamics quantities is formally obtainable, {e.g.}
\begin{equation}
C_V=T(\frac{\partial{S}}
           {\partial{T}}), \;\;
M=-(\frac{\partial{F}}{\partial{H}}), \;\; 
\chi=(\frac{\partial{M}}{\partial{H}}).
\end{equation}
For the sake of saving space, we did not write them out.

The free energy (\ref{eq:explictFE}) is conveniently partitioned as
two parts. $F=F_e+F_i$, here $F_i$ stands for the contribution
from the impurity,
\begin{eqnarray}
F_i=\frac{T}{\pi}\int\frac{u}{u^2+k^2}\ln(1+e^{-\epsilon(k)/T})dk
             \nonumber\\ 
   -\frac{T}{\pi}\sum_n\int\frac{nu/2}{(nu/2)^2\eta^2+\lambda^2}
   \ln(1+e^{-\xi_n(\lambda)/T})d\lambda, \nonumber
\end{eqnarray}
and $F_e$ for that from the electrons. 

\subsection{Strong coupling limit}

For the strong coupling $u>>1$ and non-vanishing  
external magnetic field,

we keep the leading term in eq.(\ref{eq:epsilon}) and (\ref{eq:xi}),
\begin{eqnarray}
\epsilon(k)=k^2-H-\mu,   \nonumber\\
\xi_n(\lambda)=2nH.
\label{eq:Sepsilonxi}
\end{eqnarray}
The free energy related to impurity and electrons become
\begin{eqnarray}
F_i &=& T\int\frac{dk}{\pi}\frac{u}{u^2+k^2}
          \ln[1+e^{(\eta-k^2)/T}]  \nonumber\\
 \,&\,&\, -T\sum_n\int\frac{d\lambda}{\pi}\frac{nu/2}{(nu/2)^2+\lambda^2}
             \ln[1+e^{-2nH/T}],
            \nonumber\\
F_e &=& \mu N-T\frac{L}{2\pi}\int\ln[1+e^{(\eta-k^2)/T}]dk,
\label{eq:Sfenergy}
\end{eqnarray}
where $\eta=\mu+H$.
Since $\tan^{-1}(k/u)\simeq k/u$ for $u>>1$, integrating 
eq.(\ref{eq:Sfenergy})  we have  
\begin{eqnarray}
F_i &=& \frac{4}{\pi u}\int_0^\infty\frac{k^2}{1+e^{(k^2-\eta)/T}}dk
-T\sum_n\ln(1+e^{-2n H/T}),  \nonumber\\  
F_e &=& \mu N-\frac{2L}{\pi}\int_0^\infty\frac{k^2}{1+e^{(k^2-\eta)/T}}dk.
\label{eq:Sfenergy2}
\end{eqnarray}

Now we compute the common integration in eq.(\ref{eq:Sfenergy2}),
\begin{equation}
I=\int_0^\infty\frac{k^2}{1+e^{(k^2 eta/T)}}dk,
\label{eq:I1}
\end{equation}
Changing variables of $z=(k^2 -\eta)/T$ brings it to
\begin{equation}
I=\frac{T}{2}\int_{-\eta/T}^{\infty}\frac{(zT+\eta)^{1/2}}{1+e^z}dz.
\label{eq:I2}
\end{equation}
which is conveniently slit up into three terms,
\twobeone
\begin{equation}
I=\frac{T}{2}\int_0^{\eta/T}(\eta-zT)^{1/2}dz
  -\frac{T}{2}\int_0^{\eta/T}
     \frac{(\eta-zT)^{1/2}}{1+e^z}dz
  +\frac{T}{2}\int_0^\infty
    \frac{(\eta+zT)^{1/2}}{1+e^z}dz.
\label{eq:I3}
\end{equation}
The right end of the interval for integration in the second term 
can be regarded as 
$\infty$ due to the contribution from 
large value of $z$ is negligible. 
Integrating eq. (\ref{eq:I3}) after expanding the numerator as Taylor series, 
we obtain  
\begin{equation}
I=\frac{1}{3}\eta^{3/2}+\frac{T}{2}\sum_{n=0}^\infty
      (-1)^n\frac{\Gamma(n+1/2)}{2\sqrt{\pi}}
      T^{2(n+1/2)}[1-\left({1\over 4}\right)^{n+1/2}]\zeta(2n+2)
       \left(\frac{1}{\eta}\right)^{n+1/2}
\label{eq:I4}
\end{equation}
\onebetwo
where $\zeta(2n+2)$ is the Riemann Zeta function. 
Consequently, the free energy is obtained
\begin{eqnarray}
F_i &=& \frac{4}{\pi u}I-T\sum_{n=1}^\infty\ln[1+e^{-2nH/T}],
        \nonumber\\
F_e &=& \mu N-\frac{2L}{\pi}I.
\label{eq:explictFE}
\end{eqnarray}

\subsection{Thermodynamic quantities at low temperature}
 
In the low temperature approximation, eq.(\ref{eq:explictFE}) becomes
\begin{equation}
F_i=\frac{4}{\pi u}I-T\sum_{n=1}^\infty {e}^{-2nH/T}.
\label{eq:F}
\end{equation}

{\it The  magnetization:}

The contributions of electrons and the impurity to the magnetization 
are obtained 
\begin{eqnarray}
M_e &=& \frac{2L}{\pi}I_h,  \nonumber\\ 
M_i &=& -\frac{4}{\pi u}I_h
        -2\sum_{n=1}^\infty{n}e^{-2nH/T},
\label{eq:magnetization}
\end{eqnarray}
where
\begin{equation}
I_h=\frac{\partial{I}}{\partial{H}},\;\;\;\;
I_{hh}=\frac{\partial^2{I}}{{\partial{H}}^2},\;\;\;\;
I_{tt}=\frac{\partial^2{I}}{{\partial{T}}^2}.
\label{eq:theI}
\end{equation}
Clearly, the system has spontaneous  magnetization, 
\begin{eqnarray}
M_e(H\rightarrow 0) &=& 
        \frac{L}{\pi}\mu^{1/2}-\frac{\pi L}{24}T^2\mu^{-3/2},
          \nonumber\\
M_i &=& -\frac{2}{\pi u}\mu^{1/2}+\frac{\pi}{12u}\mu^{-3/2}.
      \nonumber
\end{eqnarray}

{\it Specific heat and magnetic susceptibility}:

Specific heat and the magnetic susceptibility are 
\begin{eqnarray}
C_e &=& \frac{2LT}{\pi}I_{tt},\nonumber\\
\chi_e &=& -\frac{2L}{\pi}I_{hh},\nonumber\\
C_i &=& -\frac{4T}{\pi u}I_{tt}+\frac{4H^2}{T^2}e^{-2H/T},
   \nonumber\\
\chi_i&=&-\frac{4}{\pi u}I_{hh}+\frac{4}{T}e^{-2H/T},
\label{eq:four}
\end{eqnarray}
where $C_e$($C_i$) is the specific heat of electrons (impurity), 
and $\chi_e$($\chi_i$) is the magnetic
susceptibility of electrons (impurity).

Wilson's treatment of Kondo model 
by re-normalization-group
calculation has made it possible for the determination of proportionality
factor(``Wilson Ratio'') relating low-temperature and high-temperature 
dimensional scales. When $T\rightarrow 0$, because $e^{-2H/T}$ decrease
more rapidly than $T$, the second term of eq.(\ref{eq:four}) 
can be neglected. Then we are able to evaluate the Wilson's Ratio 
\[
R=\frac{\chi_i/\chi_e}{C_i/C_e}=1. 
\]
Furthermore, if we ignore the small term in $I_{tt}$ and $I_{hh}$, that
is we only remain the first term of them, we find that the impurity's 
contribution to specific heat at low temperature is Fermi-liquid like
\begin{equation}
C_i=-\frac{\pi}{3u}(\mu+H)^{-1/2}T,
\label{eq:Ci}
\end{equation}
and so is the magnetic susceptibility
\begin{equation}
\chi_i=-\frac{1}{\pi u}(\mu+H)^{-1/2}-\frac{\pi{T}^2}{8u}
             (\mu+H)^{-5/2}.
\label{eq:susceptibility}
\end{equation}
Obviously, the zero temperature susceptibility is of finite
\begin{equation}
\chi_i(T=0)=-\frac{1}{\pi u}(\mu+H)^{-1/2},
\label{eq:chi0}
\end{equation}
indicating that the impurity spin manifest in high temperature regime
by Curie's Law, $\chi_i\propto 1/T$, is now completely screened.
We interpret the effect as due to strong effective coupling between
impurity and the conduction electrons leading to the formation of a
local singlet, and the infra-red physics is dominated by a strong
coupling fixed point.

The electron's contributions to specific heat and magnetic 
susceptibility are
\begin{eqnarray}
C_e &=& \frac{\pi L}{6}(\mu+H)^{-1/2}T,
         \nonumber\\
\chi_e &=& \frac{L}{2\pi}(\mu+H)^{-1/2}+\frac{\pi LT^2}{16}
             (\mu+H)^{-5/2}.
\label{eq:Ce}
\end{eqnarray}

We have analyzed the thermodynamics of Kondo Model with
electronic interactions, particularly discussed the case of 
strong coupling limit extensively. 
In that case we have shown the impurity's contribution
to the specific heat and magnetic susceptibility of the system
is Fermi-liquid like and shown that in very low temperature the system
has the property of spontaneous magnetization.
 
\section*{Appendix }

Since $\epsilon$ and $\xi$ which solve eq(\ref{eq:epsilon}-\ref{eq:xi})
evidently depend on $\mu$, we should consider them as functionals of $\mu$
which is usually functions of some thermodynamic variables $x$ (such as
$T$ or $L$). The derivative of free energy (\ref{eq:ffenergy}) is given by
\begin{eqnarray}
\frac{\partial F}{\partial x}
  =\left(\frac{\partial F}{\partial x}\right)_{\mu,\epsilon,\xi}
    +N\frac{\partial\mu}{\partial x} \nonumber\\
     +\int dk\frac{L/2\pi-K_1(k)}{1+e^{\epsilon(k)/T}}
      \frac{\partial\epsilon}{\partial\mu}\frac{\partial\mu}{\partial x}
               \nonumber\\
  +\sum_n\int d\lambda\frac{K_{n/2}(\lambda)}
                           {1+e^{\xi_n(\lambda)/T}}
    \frac{\partial\xi_n}
         {\partial\mu}
    \frac{\partial \mu}
         {\partial x}.
\label{eq:partialF}
\end{eqnarray}
where $(\partial F/\partial x)_{\mu,\epsilon,\xi}$ stands for
partial derivative of $F$ with respect to the explicit variable $x$
while $\mu$, $\epsilon$, and $\xi$ are regarded as irrelevant to $x$.

The derivative of (\ref{eq:epsilon}) with respect to $\mu$ is
\begin{equation}
\frac{\partial\epsilon}{\partial\mu}=
 -1+\sum_n\int d\lambda\frac{K_{n/2}(k-\lambda)}{1+e^{\xi_n(\lambda)/T}}
   \frac{\partial\xi_n}{\partial\mu}
\label{eq:partialep}
\end{equation}
Integrating (\ref{eq:partialep}) after 
multiplying both hand sides with $\rho$, we obtain
\twobeone
\begin{eqnarray}
\int dk\frac{\partial\epsilon}{\partial\mu}\rho(k)=
  -\frac{N}{L} +\sum_n\int d\lambda
    \frac{\partial\xi_n}{\partial\mu}\sigma_n(\lambda)
  -\frac{1}{L}\sum_n\int d\lambda
    \frac{K_{n/2}(\lambda)}{1+e^{\xi_n(\lambda)/T}}
     \frac{\partial\xi_n}{\partial\mu} \nonumber\\
  +\sum_{nmq}\int\int d\lambda d\lambda'\frac{\partial\xi_n}{\partial\mu}
    \frac{A_{nmq}}{1+e^{\xi_n(\lambda)/T}}
     K_{q/2}(\lambda-\lambda')\sigma_m(\lambda').
\label{eq:rhoep}
\end{eqnarray}
In deriving to the above equation, the second equation of
 eq. (\ref{eq:distributionatT})
has been used.
We take derivative of eq.(\ref{eq:xi}), then multiply both hand
sides with $\sigma_n$ and integrate over $\lambda$. 
Summing over the subscript $n$ in what we obtained, we have
\begin{eqnarray}
\sum_n\int d\lambda\frac{\partial\xi_n}{\partial\mu}\sigma_n(\lambda)
  =\sum_n\int\int dk d\lambda
    \frac{K_{n/2}(\lambda-k)\sigma_n(\lambda)}
         {1+e^{\epsilon(k)/T}} 
             \nonumber\\
  -\sum_{nmq}\int\int d\lambda d\lambda'\frac{\partial\xi_m}{\partial\mu}
    \frac{A_{nmq}}{1+e^{\xi_m(\lambda')/T}}K_{q/2}(\lambda-\lambda')
     \sigma_n(\lambda).
\label{eq:sigmaxi}
\end{eqnarray}
With the help of the first equation of eq. (\ref{eq:distributionatT}),
eq.(\ref{eq:sigmaxi}) and eq.(\ref{eq:rhoep}) give rise to
\begin{equation}
N + \int dk\frac{L/2\pi -K_1(k)}{1+e^{\epsilon(k)/T}}
     \frac{\partial\epsilon}{\partial\mu}
  + \sum_n\int d\lambda\frac{K_{n/2}(\lambda)}{1+e^{\xi_n(\lambda)/T}}
      \frac{\partial\xi_n}{\partial\mu} = 0.
\label{eq:null}
\end{equation}
\onebetwo
Thus the complete cancellation of the last three terms 
in eq. (\ref{eq:partialF})
concludes that 
\[
\frac{\partial F}{\partial x}=
\left(\frac{\partial F}{\partial x}\right)_{\mu,\epsilon,\xi}.
\]

The work is  supported by the grants NSFC No. 19675030, 
NSFZ No. 198024, and also supported by RGC grant No. 2160089.

\end{multicols}
\end{document}